\documentclass[conference, a4paper]{IEEEtran}
\usepackage{cite}
\usepackage{amsmath,amssymb,amsfonts}
\usepackage{algorithmic}
\usepackage{graphicx}
\usepackage{float}
\usepackage{textcomp}
\usepackage{xcolor}
\usepackage{subfigure}
\usepackage{caption}
\usepackage{url}
\usepackage{algorithm}
\usepackage{algorithmic}
\usepackage{amsmath}
\usepackage{algorithmic} 
\usepackage{url}
\usepackage{fancyhdr}
\usepackage{booktabs}
\def\BibTeX{{\rm B\kern-.05em{\sc i\kern-.025em b}\kern-.08em
    T\kern-.1667em\lower.7ex\hbox{E}\kern-.125emX}}
\usepackage[left=0.57in,right=0.57in,top=0.7in,includefoot,bottom=1 in]{geometry}
\begin{document}
\title{UAV-assisted Semantic Communication with Hybrid Action Reinforcement Learning}
\author{\IEEEauthorblockN{Peiyuan Si$^1$,
Jun Zhao$^1$, 
Kwok-Yan Lam$^1$,
Qing Yang$^2$}
\IEEEauthorblockA{\\$^1$Nanyang Technological University, Singapore\\
$^2$University of North Texas, the United States\\
\{peiyuan001, junzhao, kwokyan.lam\}@ntu.edu.sg, qing.yang@unt.edu}}

\maketitle
 \thispagestyle{fancy}
\pagestyle{fancy}
\lhead{This paper appears in IEEE Global Communications Conference (GLOBECOM) 2023.}
\cfoot{\thepage}
\renewcommand{\headrulewidth}{0.4pt}
\renewcommand{\footrulewidth}{0pt}
\begin{abstract}
In this paper, {we aim to explore the use of uplink semantic communications with the assistance of UAV in order to improve data collection effiicency for metaverse users in remote areas.}
To reduce the time for uplink data collection while balancing the trade-off between reconstruction quality and computational energy cost, we propose a hybrid action reinforcement learning (RL) framework to make decisions on semantic model scale, channel allocation, transmission power, and UAV trajectory. The variables are classified into discrete type and continuous type, which are optimized by two different RL agents to generate the combined action.
{Simulation} results indicate that the proposed hybrid action reinforcement learning framework can effectively improve the efficiency of uplink semantic data collection under different parameter settings and outperforms the benchmark scenarios.
\end{abstract}


\section{Introduction}

The proposal of Metaverse has been promoted by the implementation of 5G communication technology and maturing virtual reality (VR) devices in recent years \cite{AllYouNeedToKnow}. To keep the Metaverse up-to-date, {uplink data collection for} object modeling and updating are essential for VR applications. 
{The efficiency of data transmission has a direct impact on user experience once there are demands to update the VR background, which is different from the traditional VR applications whose contents are not frequently updated.}
The 3-D modeling of remote area VR backgrounds including buildings (indoor and outdoor), roads, and natural environments are based on numerous photos taken on location, e.g., more than 1500 images with the average size of 10Mb are required to model an area with historic buildings \cite{3dsurvey}. The data collection with such large size poses requirements for both high transmission efficiency and wide network coverage. Due to the limited coverage area of base stations and the randomness of update demand, it is expensive to deploy stable wireless network coverage in remote areas. Thus, an efficient and economical way for remote area image data collection is urgently needed.

To overcome the challenge of data transmission efficiency, semantic communication is proposed as another way {other than the traditional communication research} to improve communication efficiency \cite{semantic_luo2022}. Traditional communication research is ``data-oriented", which focuses on how to transmit more data bits in unit time \cite{li2018spectrum,li2019spectral}. Semantic communication focuses on the higher level of communication efficiency, i.e., to improve the semantic meaning transmission efficiency. The current study on the semantic encoder and decoder are mainly based on variational autoencoders (VAEs) \cite{xie2022task}, which encode input data into a lower-dimensional latent space and decode it back to reconstruct the original input data \cite{VAE}. In the implementation, the encoder and decoder are deployed at the transmitter and receiver respectively, and the latent space is transmitted through the data link. Although the early VAEs are only able to recover small images such as MNIST (image size $28\times28$) and CIFAR-10 (image size $32\times 32$), the state-of-art VAE models have shown great potential in semantic communication. Oord et al. \cite{VQ_VAE} published VQ-VAE in 2017, which incorporates the vector quantization technique into VAE and is able to reconstruct images and videos. Razavi et al. \cite{VQ_VAE_2} made improved the traditional VAE by utilizing hierarchical multi-scale latent maps, and their experiments show good reconstruct quality on high-quality images such as FFHQ dataset (image size $1024 \times 1024$). Nouveau VAE (NVAE) proposed by Vahdat et al. \cite{NVAE} further improved the performance of VAE and achieved satisfying results on various high-quality image datasets.
{Li et al. \cite{FAST} found that devices can select different scales of sub-models that requires less computational energy at the cost of reconstruction quality, and formulated the relationship between them. } 

To cope with the challenge of wireless network coverage in remote areas, UAV-assisted data collection is considered as a practical solution to set up flexible wireless networks for heterogeneous user requirements \cite{RuiZhangUAV}, {especially the research on UAV-enabled communication resource allocation, trajectory control, and inter-UAV cooperation} \cite{SiIoTJ}. 
UAV-based optimization problems which take the trajectory into consideration usually segment the flight time of UAV into certain numbers of discrete time slots for the convenience of computation. In such case, a set of resource allocation variables need to be optimized in each time slot to ensure optimal performance.
If the flight time of the UAV in the system model is uncertain, the optimization problem becomes a time-sequential problem that is hard to solve because the number of variables can not be determined.

To solve the time-sequential UAV trajectory-related problems, reinforcement learning (RL) has attracted much research interest due to its ability in maximizing long-term reward \cite{li2022dynamic}.
Cui et al. \cite{Cui_MultiUAV} proposed a multi-agent reinforcement learning resource allocation algorithm for multi-UAV networks and showed fast convergence with the basic Q-learning algorithm. 
Luong et al. \cite{luong2021deep} utilized the deep Q-learning algorithm to learn the network state for the decision of the movement of UAV and improved the network performance by up to 70\%. 
Rodriguez-Ramos et al. \cite{UAVland} implemented a versatile Gazebo-based reinforcement learning framework for UAV landing on a moving platform, which is a novel experiment of deep deterministic policy gradient (DDPG) on UAV controlling.

{In this paper, we aim to minimize the time for the semantic data collection mission in remote area and to balance the reconstruction quality and energy cost for computation by reinforcement learning approach.
Traditional reinforcement learning algorithms that focus on either discrete action or continuous action can not solve the hybrid action communication problem that we focus on, i.e., there are both discrete action and continuous actions. To this end, we propose a hybrid action reinforcement learning framework with two cooperative agents for discrete action (channel allocation) and continuous actions (transmission power, semantic model scale, and UAV trajectory), respectively.
To model the relationship between reconstruction quality and semantic model scale, we adopt a fitted logarithmic function according to the latest study \cite{FAST}. For generality, we set tunable importance factors to denote the weight of reconstruction quality and computational energy cost so that the system can work in different modes.}

The contributions of this paper are as follows:\begin{itemize}
\item We propose a hybrid action reinforcement learning framework for UAV-assisted semantic data collection to minimize the data collection time by optimizing channel selection, transmission power and UAV trajectory.
\item We propose a utility function with flexible importance weights to balance the reconstruction quality of semantic data and the computational energy cost. We include the semantic model scale into the action space of reinforcement learning agent according to its affect on reconstruction quality revealed by the latest research \cite{FAST}.
\item {Simulation results} indicate that the proposed framework can better reduce the data collection efficiency while balancing the reconstruction quality and computational energy cost under different parameter settings compared to the benchmark scenarios.
\end{itemize}

The rest of this paper is organized as follows. Section II introduces the system model. The problem formulation and the hybrid action reinforcement learning framework are presented in Section III and Section IV, respectively. Section V shows the evaluation of simulation results. The conclusion of the paper is discussed in Section VI.

\section{System Model}
\begin{figure}[htb]
 \centering
 \includegraphics[width=0.96\linewidth]{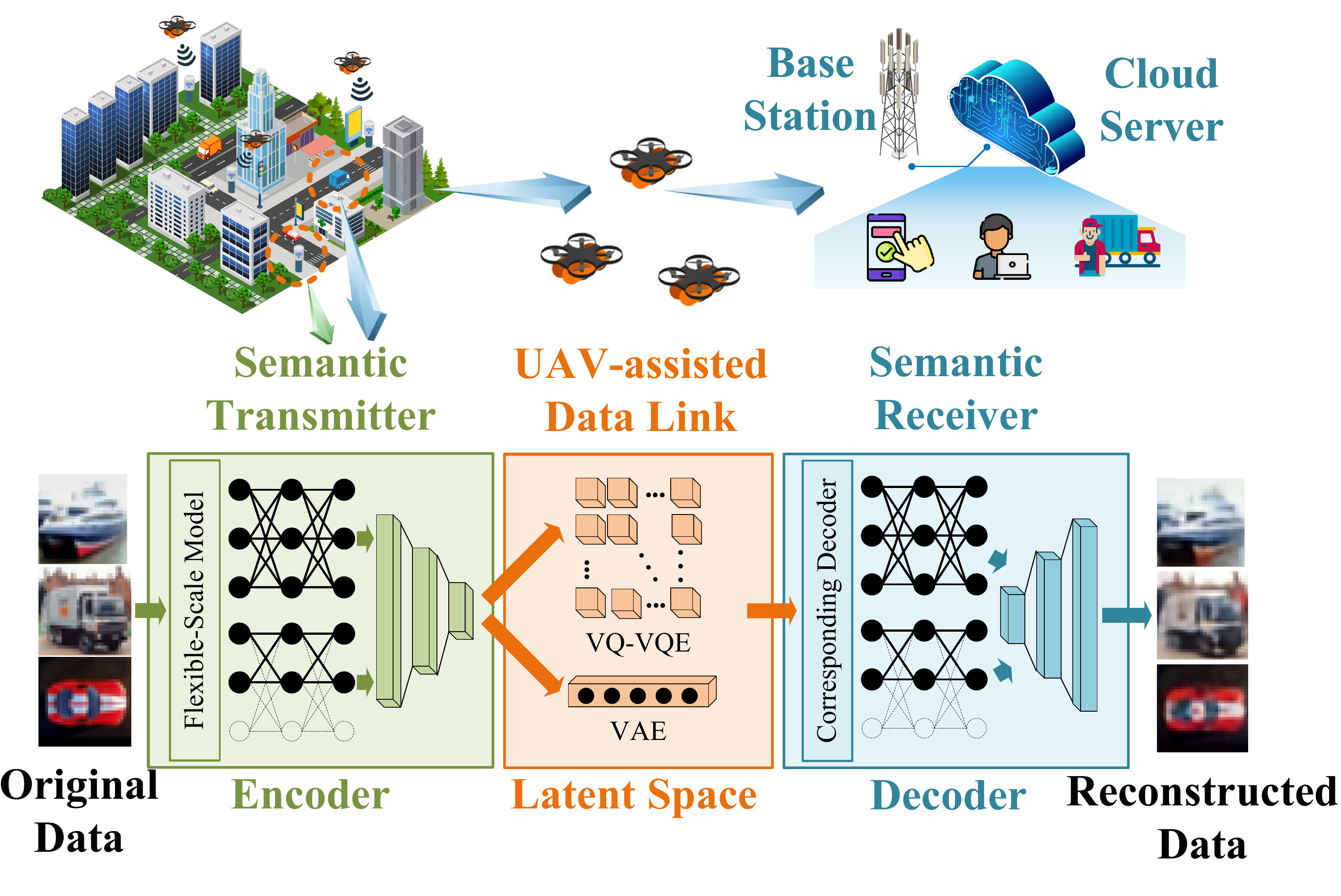}
 \caption{UAV-assisted semantic data collection in remote area.}
 \label{fig:System_Model}
\end{figure}
As shown in Fig. \ref{fig:System_Model}, $N$ mobile users with uploading demand of collected remote background data for VR modeling are located in a given $L\times L$ remote area. The 3-D location of the $n$-th user is denoted by$(x_n,y_n,0)$. The UAV trajectory is discretized into $T$ time slots with equal length $t_\text{slot}$ , and the UAV is assumed to be stationary in each time slot. The UAV for data collection flies at fixed height $H$, and its location at time slot $t$ is denoted by $(x_{\text{uav}}[t], y_{\text{uav}}[t], H)$, where $H$ denotes the height.
To improve the uploading efficiency, the original data is encoded into the latent space by personal mobile devices, and is collected by the UAV through $M$ orthogonal channels. Once the UAV returns to the base station, the collected data is uploaded to the cloud server for reconstruction. To simplify the model and focus on the main problem, we assume that each user has the same datasize $U$ to be uploaded. \subsection{Reconstruction Quality and Computational Energy Cost}
\begin{figure}[h]
 \centering
 \includegraphics[width=0.9\linewidth]{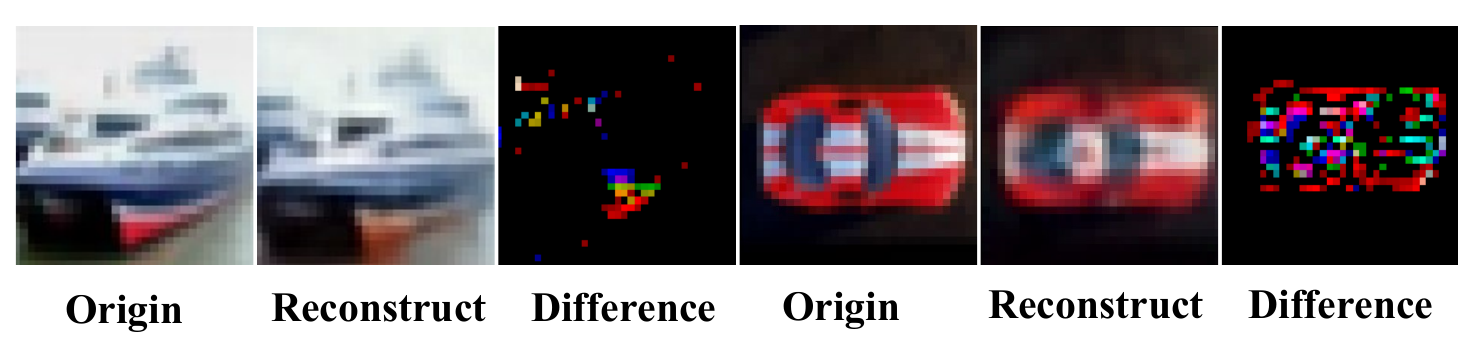}
 \caption{Difference between original images in CIFAR-10 dataset and images reconstructed by VQ-VAE-2 model.}
 \label{diff}
\end{figure}

Due to the limitation in computational resources and energy of personal devices, we adopt the flexible-scale model for the encoder and the decoder to achieve the balance between reconstruction quality and computational energy cost. The reconstruction quality $Q(\eta)$ is defined by \cite{FAST}
\begin{align}
Q(\eta )=1-\frac{1}{D\left| {{X}_{\text{test}}} \right|}\sum\nolimits_{x\in {{X  }_\text{test}}}{\left\| x-F\text{(}x,\theta ,\eta \text{)} \right\|},
\end{align}
{where $\eta \in (0,1]$ denotes the scale ratio of the sub-model compared to the full-size model}, $D$ denotes the {number of pixels in the image}, $F(x,\theta,\eta)$ denotes {the pixel value matrix} of the reconstructed image, $\theta$ denotes the weights in the model, $x$ denotes {the pixel value matrix of} the original image, $\left| {{X}_{\text{test}}} \right|$ denotes the number of samples in the test dataset $X_\text{test}$, and $\left\| \cdot\right\|$ denotes the L1 distance between the original image and the reconstructed image. Fig. \ref{diff} shows visualized difference between original images in CIFAR-10 dataset and images reconstructed by VQ-VAE-2 model \cite{VQ_VAE_2}.
The latest research revealed the relationship between $Q(\eta)$ and $\eta$ through extensive experiments, {which is given by}
\begin{align}
    Q(\eta )={{\omega }_{1}}\ln \left( \frac{{{\omega }_{2}}}{\eta }+{{\omega }_{3}} \right)+{{\omega }_{4}},
\end{align}
where ${\omega }_{1},{\omega }_{2},{\omega }_{3},{\omega }_{4}$ are hyper-parameters to fit the data obtained from experiments \cite{FAST}.
\subsection{Channel Setting and Transmission Rate Calculation.}
In the proposed data collection model, each user can occupy only one channel, but multiple users are able to share the same channel through non-orthogonal multiple access (NOMA).
The number of users in channel $m$ is denoted as $N_m$.
According to the experimental characterization of the vehicle-to-infrastructure radio channels in suburban environments implemented by Yusuf et al \cite{channel}, the small-scale fading of the strongest path is found to be Rician distributed. The channel gain matrix $\mathbf{h}[t]$ contains all the channel gain information between UAV and users at $t$-th time slot, whose elements are given by \cite{3DTraj}
\begin{align}
{{h}_{n,m}}[t]=\sqrt{{{\beta }_{n}}[t]}{{g}_{n,m}}[t],
\label{h}
\end{align}
where $n$ and $m$ denotes the index of user and channel, respectively. ${{\beta }_{n}}[t]$ denotes the large-scale average channel gain at time slot $t$, and ${{g}_{n,m}}[t]$ denotes the small-scale fading coefficient, which is modelled as Rician fading. ${{\beta }_{n}}[t]$ and ${{g}_{n,m}}[t]$ are given by
\begin{align}
{{\beta }_{n}}[t]={{\beta }_{0}}d_{n}^{-\alpha }[t],
\label{beta}
\end{align}
and
\begin{align}
{{g}_{n,m}}[t]=\sqrt{\frac{K}{K+1}}g+\sqrt{\frac{1}{K+1}}\tilde{g},
\label{g}
\end{align}
where $\beta_0$ denotes  the channel gain at the reference distance ${{d}_{0}}=1$m, $\alpha$ denotes the path loss exponent, which varies from 2 to 6 (in this paper we assume that $\alpha=2$). $g$ denotes the deterministic LoS channel component with $|g|=1$, which denotes the randomly scattered component. The Rician factor is denoted by $K$. $d_{n}[t]$ denotes the distance from UAV to user $n$ in time slot $t$, which is given by
\begin{align}
d_{n}^{{}}[t]=\sqrt{{{({{x}_{n}}-{{x}_{\text{uav}}}[t])}^{2}}+{{({{y}_{n}}-{{y}_{\text{uav}}}[t])}^{2}}+{{H}^{2}}}.
\label{d}
\end{align}
The channel-to-noise-ratio (CNR) is given by
\begin{align}
    {{\Gamma }_{n,m}}[t]=\frac{{{h}_{n,m}}[t]}{{{B\sigma }^{2}}}
\end{align}
where $\sigma^2$ denotes the power spectral density of additive white Gaussian noise (AWGN) at the receiver, and $B$ denotes the bandwidth.
The signal to interference plus noise ratio (SINR) of user $n$ in channel $m$ in time slot $t$ is given by
\begin{align}
{{\gamma }_{n,m}}[t]=\frac{{{p}_{n}}[t]{{\Gamma }_{n,m}}[t]}{1+\sum\limits_{i={{J}_{n,m}}+1}^{{{N}_{m}}}{{{p}_{{{n}_{i}}}}[t]{{\Gamma }_{{{n}_{i}},m}}[t]}}.
\end{align}
where $p_{n}$ denotes the user transmission power, {$J_{n,m}$ denotes the rank of transmission power of user $n$ in channel $m$.}
Thus, the transmission rate of user $n$ in channel $m$ and time slot $t$ is given by
\begin{align}
    {{R}_{n,m}}[t]=B{{\log }_{2}}(1+{\gamma }_{n,m}[t]).
\end{align}

\section{Problem Formulation}
Our goal is to minimize the required time $T$ for UAV to finish the semantic data collection mission, and to balance the semantic data reconstruction quality and computational energy cost. 
The objective of the RL agents is to find a set of discrete action $a_t^d$ for channel selection and continuous action $a_t^c$ (consists of transmission power $a_t^p$, model scale $a_t^{\eta}$ and UAV trajectory $a_t^x$ and $a_t^y$) that achieves
\begin{align}
    &({{a}_{c}},{{a}_{d}})=\arg \max (\hat{U}-T).\nonumber\\
    \text{subject to }&\sum\nolimits_{n\in N,m\in M,t\in T}{{{R}_{n,m}}[t]}\ge N\cdot U,
\end{align}where $U$ denotes the data size of each user, $N$ denotes the number of users. Constraint (10) ensures the completion of data collection from all the users. $\hat{U}$ denotes the utility function that considers computational energy cost and reconstruction quality, which is given by
\begin{align}
   \hat{U}=\lambda \sum\nolimits_{n\in N}Q(\eta )-(1-\lambda )\sum\nolimits_{n\in N}{{{E}_{n}}},
\end{align}
where $\lambda$ is an adjustable importance factor of reconstruction quality, e.g., the model works at quality-first mode if $\lambda=1$, and works at energy-efficient mode if $\lambda=0$. $E_n$ denotes the computational energy cost, which is given by \cite{FAST}
\begin{align}
    {{E}_{n}}=\mathcal{K}{{\eta }_{n}}^{2}\left( {{\varepsilon }_{e,n}}f_{e,n}^{2}{{W}_{e}}+{{\varepsilon }_{d,n}}f_{d,n}^{2}{{W}_{d}} \right),
\end{align}
where $\mathcal{K}$ denotes the size of latent space, $\eta_n$ denotes the model scale ratio of the $n$-th user. ${\varepsilon }_{e,n}$ and ${\varepsilon }_{d,n}$ denote the hardware energy coefficients encoder and decoder, respectively. $f_{e,n}$ and $f_{d,n}$ denote the computing frequency of encoder and decoder, respectively. ${{W}_{e}}$ and ${{W}_{c}}$ denote the computing workload at the encoder and decoder, respectively.

\section{Hybrid Action Generation Model}
\begin{figure}[htb]
 \centering
 \includegraphics[width=0.9\linewidth]{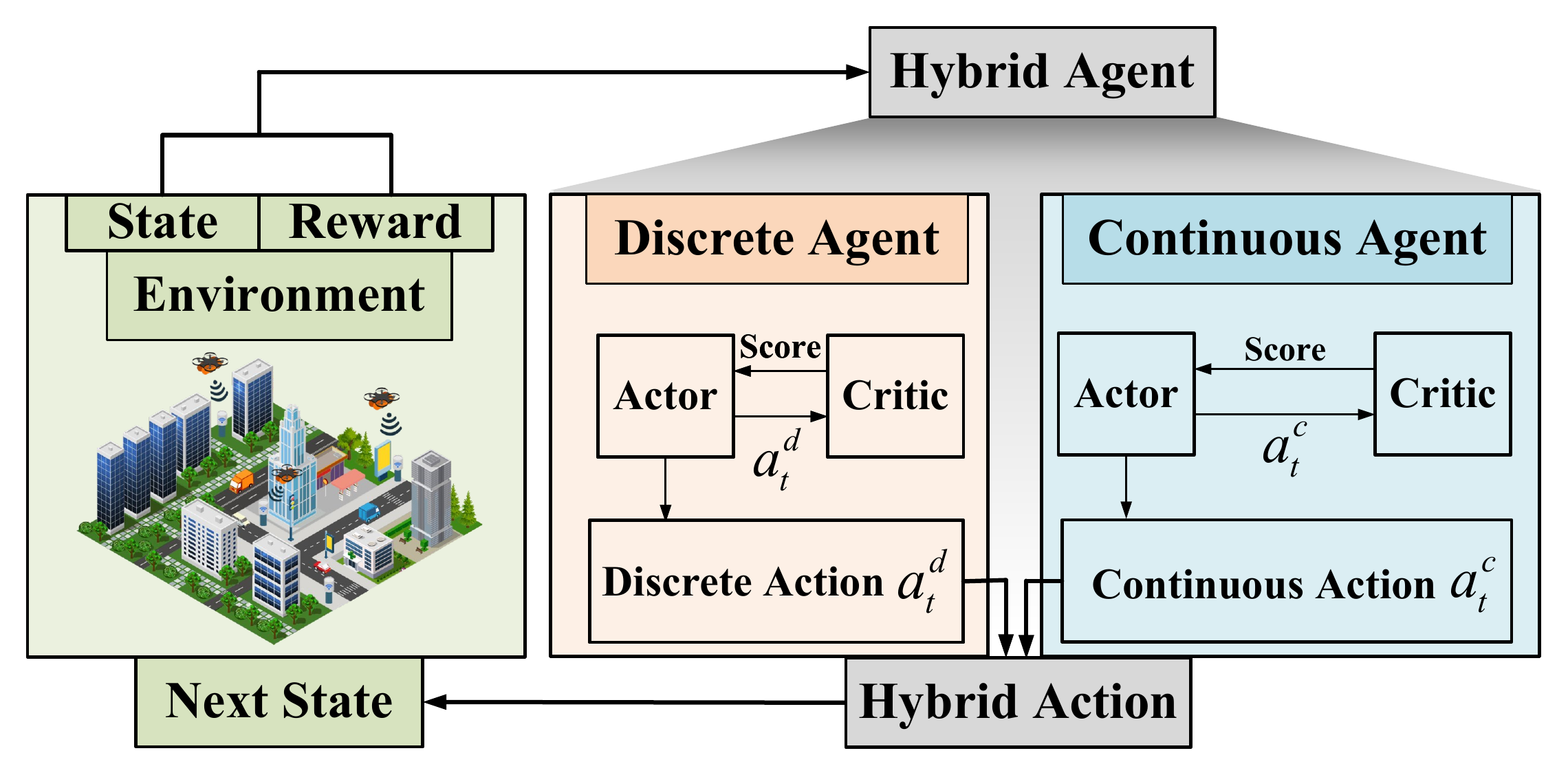}
 \caption{Double-agent policy generation model.}
 \label{solution}
\end{figure}

 As shown in Fig. \ref{solution}, we introduce the hybrid action generation model to minimize the required time for semantic data collection while balancing the semantic model scale and computational energy cost.
In the proposed model, actions are classified into discrete action $a_d$ and continuous action $a_c$. The hybrid agent consists of two proximal policy optimization (PPO) \cite{PPO_origin} based sub-agents with discrete action space and continuous action space, respectively. Given the current state, two sub-agents work cooperatively to generate the hybrid action to update the next state of the environment and receive the corresponding reward.

\subsection{Discrete Agent for Channel Selection}
The discrete agent for channel selection consists of a critic network and an actor network with discrete action space. 
The critic network is responsible to give scores to the actor according to its output action. The loss function of the critic network $J^{d}(\phi )$ is given by
\begin{align}
J^{{d}}(\phi )={{\left[ V_{\phi }^{{d}}(s_{t}^{{d}})-\left( r_{t}^{d}+\gamma V_{{{\phi }'}}^{d}(s_{t+1}^{{d}}) \right) \right]}^{2}},
\label{a_loss_dis}
\end{align}
where $V_{\phi }^{d}(s_{t}^{d})$ and $V_{{{\phi }'}}^{d}(s_{t+1}^{d})$ denote the state value estimations of the current and next state, respectively. $\phi$ and $\phi'$ denote the model parameters in the current and next step, respectively. $r_t^{d}$ denotes the reward, and $s_t^{d}$ denotes the state in the $t$-th time slot.

The actor network is similar to a classifier, where the input states go through fully connected layers and a softmax layer to generate a set of probabilities related to the action set. The agent sample from the discrete distribution to get the final action decision. The action, state, and reward are given by:

\textbf{Action:}
The index of the selected channel for the $n$-th user is denoted by $\hat{I}[t]\in [0,M]$. The empty selection ``0'' means that the user is not allocated to any channel. The encoded action is given by
$\sum\nolimits_{n\in N}{{{{\hat{I}}}_{n}}[t]{{M}^{n-1}}}$. During the interaction with the environment, the obtained action is decoded through the reverse operation to get the actual channel selection. 

\textbf{State:}
The state of the discrete agent $S_{t}^{d}=\{{{U}_{\text{res}}}[t],\mathbf{h}[t]\}$ includes the channel gain matrix $\mathbf{h}[t]$ and the remaining data at users ${U}_{\text{res}}[t]$ at $t$-th step. The remaining data is updated by
\begin{align}
    {{U}_\text{res}}[t+1]={{U}_\text{res}}[t]-{{t}_\text{slot}}\sum\nolimits_{n\in N,m\in M}{{{R}_{n,m}}[t]}.
\end{align}

\textbf{Reward:}
The reward $r_t^d$ of discrete agent consist of a penalty $r_{t}^{\text{time}}$ which is {negatively correlated to data collection time} and an utility-based reward $r_t^u$ which is given by
\begin{align}
    r_{t}^{u}=\lambda \sum\nolimits_{n\in N} Q(\eta )-(1-\lambda )\sum\nolimits_{n\in N}{{{E}_{n}}}.
\end{align}
An additional failure penalty $r_{t}^{\text{fail}}$ with {large negative value} will be given to the agent if it fails to finish the mission in the given time limit $T_{\text{max}}$. The reward of the discrete agent is given by
\begin{equation}
r_{t}^{\text{d}}=\left\{ \begin{aligned}
  & r_{t}^{\text{time}}+r_{t}^{u},\text{if }t\le {{T}_{\max }} \\ 
 & r_{t}^{\text{time}}+r_{t}^{u}+r_{t}^{\text{fail}},\text{ }\text{if }t>{{T}_{\max }} \\ 
\end{aligned} \right.
\end{equation}

\subsection{Continuous Agent for Multiple Heterogeneous Variables}
The continuous agent is responsible for semantic model scale, transmission power, and UAV trajectory, which are heterogeneous in physical meanings and units. To include them in a single agent, normalization is necessary before forwarding them to the neural networks. The structure of the critic network of the continuous agent is the same as the discrete one, but there are differences in the actor network due to the continuous nature of the action space. The output of the actor network has two heads: mean head $\mu$ and variance head $\sigma$. A set of Gaussian distributions $\mathcal{N}(\mu, \sigma^2)$ are determined by the set of mean and variance, from which the output actions are sampled.
The loss function of the critic network is given by
\begin{align}
J^{c}(\phi )={{\left[ V_{\phi }^{c}(s_{t}^{c})-\left( r_{t}^{c}+\gamma V_{{{\phi }'}}^{c}(s_{t+1}^{c}) \right) \right]}^{2}}.
\label{a_loss_con}
\end{align}
The action, state and reward are designed as
follows:

\textbf{Action:}
The action of the continuous agent $a_t^{c}$ consists of three parts, which is given by
\begin{align}
    a_t^{c} = \{a_t^{\eta}, a_t^p, a_t^x,a_t^y\},
\end{align}
where $a_t^{\eta}$ denotes semantic model scale, $a_t^p$ denotes transmission power, and $a_t^x,a_t^y$ denote the movement of the UAV in the $t$-th time slot. Due to the limitation of the UAV speed, there is a constraint on the trajectory action
\begin{align}
a_{t}^{x},a_{t}^{y}\in \left[ -{{t}_{\text{slot}}}{{V}_{\max }},{{t}_{\text{slot}}}{{V}_{\max }} \right],
\end{align}
where $t_\text{slot}$ denotes the length of a single time slot, and $V_\text{max}$ denotes the maximum speed of UAV. The scale of $a_t^{\eta}$ and $a_t^p$ are constrained by $a_t^{\eta} \in [0,1]$ and $a_t^p \in [0, P_\text{max}]$, where $P_\text{max}$ is the maximum transmission power. Three types of actions with heterogeneous feasible regions are normalized during the training process, and recovered to true value during the interaction with the environment.

\textbf{State:}
The state of the continuous agent is similar to the discrete agent, which includes the current channel gain $\mathbf{h}[t]$ matrix and remaining data at users ${{U}_{\text{res}}}[t]$. In addition, the current horizontal location of UAV $\left( {{x}_{\text{uav}}}[t],{{y}_{\text{uav}}}[t] \right)$ is also included in the state $S_{t}^{\text{traj}}$, which is given by
$S_{t}^{c}=\left\{ {{U}_{\text{res}}}[t],\mathbf{h}[t], {{x}_{\text{uav}}}[t],{{y}_{\text{uav}}}[t] \right\}$.

\textbf{Reward:}
The reward of the continuous agent is modified based on the reward for the discrete agent $r_{t}^{d}$. We give an penalty $r_t^\text{penalty}$ with {fixed large negative value} to the agent if UAV goes beyond the $L \times L$ mission area to regularize the trajectory decision. The reward of the continuous agent is given by
\begin{equation}
r_{t}^{c}=\left\{ \begin{aligned}
  & r_{t}^{d},\text{ }\text{if }{{x}_{\text{uav}}}[t]\in [-L,L],{{y}_{\text{uav}}}[t]\in [-L,L] \\ 
 & r_{t}^{d} + r_{t}^{\text{penalty}},\text{ otherwise.} 
\end{aligned} \right.
\end{equation}

\section{Simulation Results}
\subsection{{Simulation} Settings}
In the simulation, we test the performance of our proposed method (\textbf{Hybrid Action}) together with the following benchmarks:
(1) \textbf{Equal Power Allocation (EP)} set the transmission power of users equally, which is the only difference from our proposed method (Hybrid Action);
(2) \textbf{Triple PPO} uses an independent agent for power control while keeping the other parts the same as Hybrid Action. The agents in our proposed method and the benchmarks are based on PPO, so there are three PPO-based agents in the benchmark triple PPO.
The hyper-parameters for computation energy $\{W_e, W_d, \epsilon_e, \epsilon_d, \mathcal{K}\}$ are empirically set as \{0.65 MCycles, 3.25 MCycles, $1^{-26}$, $1^{-26}$, 512\}. 
The hyper-parameters to fit the reconstruction quality $\{\omega_1,\omega_2, \omega_3, \omega_4\}$ are given by $\{-0.0815, 10.7192, -0.7957, 1.0918\}$ \cite{FAST}.
The other settings are given in Table \ref{table:parameter}.

\begin{table}[tbp]
\caption{Experiment Parameter Setting} \label{table:parameter}
\begin{center}
\begin{tabular}{c c}
\toprule[1pt]
 Parameter and Physical Meaning            & Value                  \\ \hline
  Number of channels($M$)                  & $3$                    \\
  Default number of users ($N$)            & $5$                    \\
  Bandwidth ($B$)                          & $5$MHz                 \\
  Transmission power of users               & $5$W                   \\
  Frequency ($f$)                          & $28$GHz (5G spectrum)  \\
  Power of Gaussian noise ($\sigma^2$)     & $5\times {{10}^{-8}}$W \\
  Maximum speed of UAV                     & $10$m/s                \\
  Mission area size ($L$)                  & $200$m                 \\               
\bottomrule[1pt]
\end{tabular}
\end{center}
\end{table}

\subsection{Performance Evalutions}
\begin{figure}[htb]
 \centering
 \includegraphics[width=0.9\linewidth]{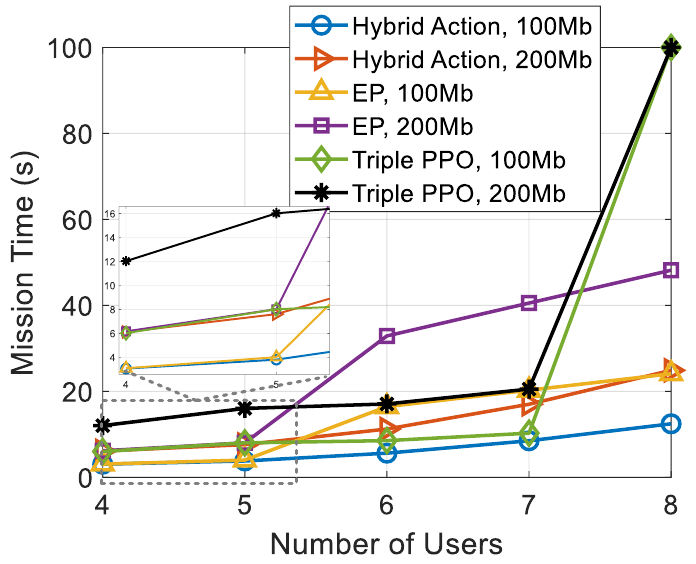}
 \caption{Required data collection mission time with different number of users.}
 \label{fig:UserNum}
\end{figure}
The comparison of required data collection mission time with different number of users are shown in Fig. \ref{fig:UserNum}. With the increase of the number of users, all of the three methods show increasing tendencies of data collection mission time, but the increasing rates are different. Our proposed Hybrid Action algorithm requires the least mission time regardless of the user number and data size. The EP method achieves similar performance as Hybrid Action when there are small number of users (four or five) because the transmission rate is degraded much by interference.  The mission time of EP increases faster than Hybrid Action when the number of users exceeds six due to the missing power control, which is critical in reducing interference. We can also find in the zoomed-in figure that performance difference starts to appear at user number five. The triple PPO method performs worse than Hybrid Action, and its mission time increases rapidly at user number eight, which indicates that the concatenating of agents degrades the system stability.
\begin{figure}[htb]
 \centering
 \includegraphics[width=0.9\linewidth]{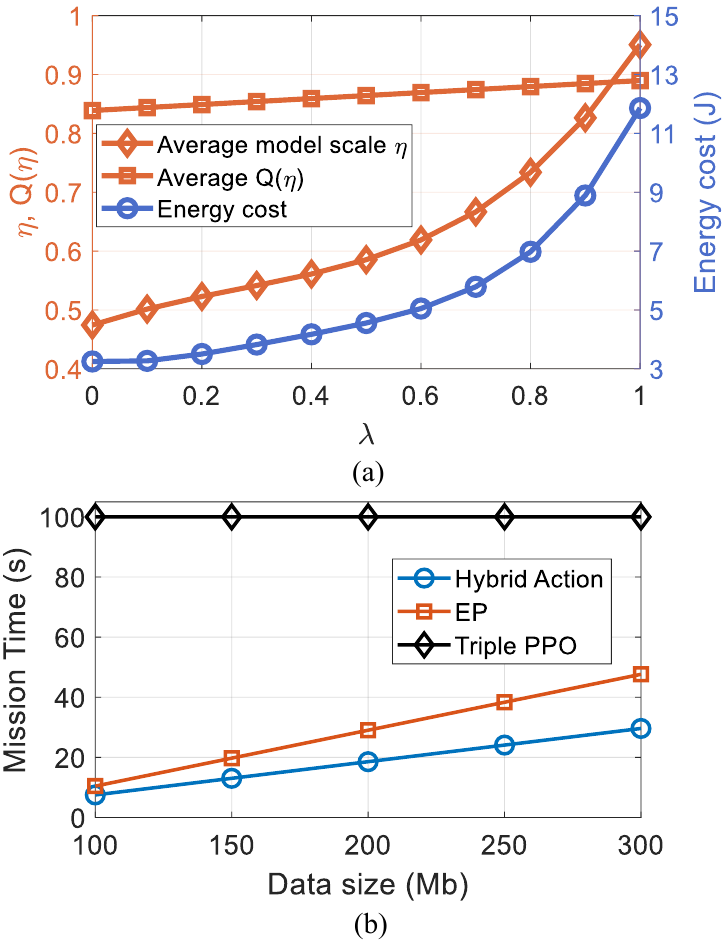}
 \caption{(a): semantic model scale, reconstruction quality and energy cost with different $\lambda$; (b): semantic data collection mission time vs data size.}
 \label{fig:lamda+datasize}
\end{figure}

Fig. \ref{fig:lamda+datasize} (a) presents the semantic model scale $\eta$, reconstruction quality $Q(\eta)$ (left y-axis), and energy cost (right y-axis) with different $\lambda$. As $\lambda$ denotes the importance of $Q(\eta)$, larger $\lambda$ leads to better reconstruction quality and a larger model scale. To support the increasing computation workload, the energy cost also increases with respect to $\lambda$. Fig. \ref{fig:lamda+datasize} (b) presents the required mission time with different data sizes at users (number of users $N=8$, $\lambda=0.5$). We can find that there is a linear relationship between data size and mission time of EP and Hybrid Action, indicating that the efficiency of our proposed method is not influenced by the workload. The mission time of the triple PPO reaches the maximum value $T_\text{max}=100$s because it can not generate reasonable actions under $N=8$ due to its unstable structure.

\begin{figure}[htb]
 \centering
 \includegraphics[width=0.9\linewidth]{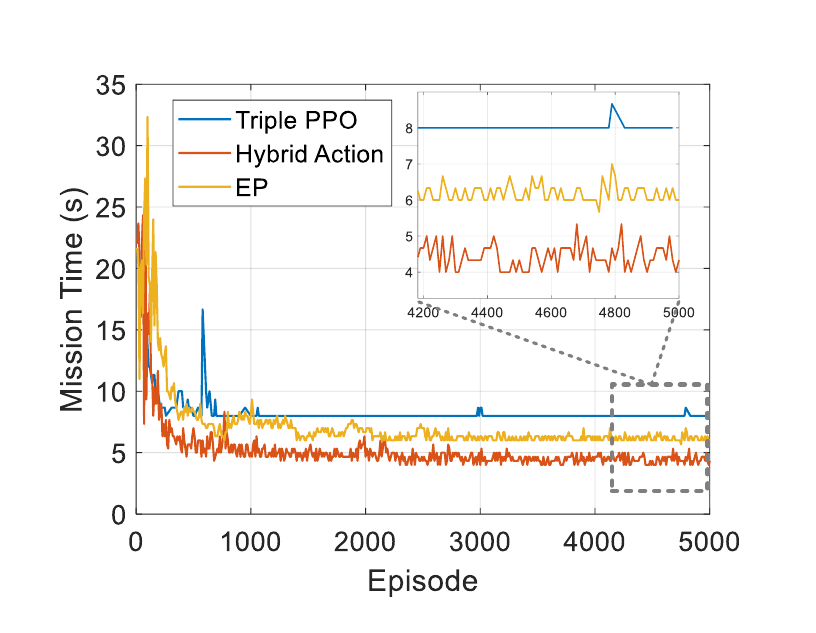}
 \caption{Training process of the proposed and benchmark RL algorithms.}
 \label{fig:Convergence}
\end{figure}
Fig. \ref{fig:Convergence} shows the training process of the algorithms involved in the experiment. With the parameter setting  $N=5$ and data size $100$Mb, all of the three algorithms reach convergence within 5000 episodes. We can observe from the zoomed-in figure that there are slight fluctuations of the mission times of the Hybrid Action and EP algorithm in the last 1000 episodes, while the triple PPO algorithm is more stable. The small fluctuation is caused by the randomness of channel gain in each episode, which directly influence the transmission rate. The randomness has less impact on the triple PPO algorithm because it generates sub-optimal actions that always require more mission time than EP and Hybrid Action. Fig. \ref{fig:UserNum}, \ref{fig:lamda+datasize}, and \ref{fig:Convergence} show that the proposed Hybrid Action algorithm outperforms the benchmarks under different parameter settings in semantic data collection missions, and its RL model structure is more robust than the benchmark triple PPO algorithm.

\section{Conclusion}
In this paper, we propose a double-agent hybrid action reinforcement framework for UAV-assisted semantic data collection. Two agents are responsible to make decisions on discrete and continuous actions, respectively. The final action consists of decisions on channel allocation, UAV trajectory, transmission power, and semantic model scale is generated to update the state of the environment. By setting the importance factor, the system can be generalized to different modes. Experiments indicate the advantage of the proposed framework over two benchmarks in both performance and robustness.


\end{document}